# Determination of Carrier Type Doped from Metal Contacts to Graphene by Channel-Length-Dependent Shift of Charge Neutrality Points

Ryo Nouchi[1*], Tatsuya Saito[2], and Katsumi Tanigaki[1,2]

[1]*WPI Advanced Institute for Materials Research, Tohoku University, Sendai 980-8578, Japan*

[2]*Department of Physics, Graduate School of Science, Tohoku University, Sendai 980-8578, Japan*

A method for determining the type of charge carrier, electron or hole, which is transferred from metal contacts to graphene, is described. The Dirac point is found to shift toward more negative (positive) gate voltages for electron (hole) doping by shortening of the interelectrode spacing. The shift of the Dirac point is accompanied by an enhancement of the electron-hole conductivity asymmetry. Experimentally determined carrier types may be explained in terms of the metal work functions modified by interactions with graphene.

* E-mail address: nouchi@sspns.phys.tohoku.ac.jp

Graphene exhibits ultrahigh charge-carrier mobilities of more than 200,000 cm$^2$ V$^{-1}$ s$^{-1}$ for a suspended structure[1] and even 40,000 cm$^2$ V$^{-1}$ s$^{-1}$ on a SiO$_2$ substrate at room temperature.[2] In order to exploit such high mobility, an electric current through graphene is usually measured using metal electrodes; however, the resulting metal-graphene contacts are known to affect the electronic transport properties of graphene. For example, the electrostatic potential at a metal contact cannot be controlled by external electric fields such as applied gate voltage, while the potential at points sufficiently distant from the contacts can be effectively tuned by the electric field.[3] In addition, the transition distance from the "pinned" points at the contacts to the "free" regions far from the contacts has been reported to be *ca*. 1 μm.[3] This invasive nature of metal contacts makes it difficult to measure the intrinsic conductivity of graphene, even with the four-terminal configuration usually employed to exclude contact resistance.[4] The long-range potential variation leads to an asymmetry of the transfer characteristics (gate-voltage dependence of the drain current $I_D$) of graphene field-effect transistors (FETs).[5] If an easily oxidizable metal is employed as the electrode material, then the thin oxide layer formed at the metal-graphene interface can depin the charge density at the contacts.[6] The long-range variation of the electrostatic potential can also alter the shape of the transfer characteristics without charge-density pinning, which accounts for the large distortion in the experimentally observed transfer characteristics.[7,8] Metal contacts can govern the electronic transport properties of graphene, especially for short interelectrode spacings.

One considerable effect of metal contacts is the charge transfer between the metal and graphene, and the charge transfer carrier type can be determined from simple electronic transport measurements: an asymmetry is known to appear in the transfer characteristics,[5] and electron (hole) doping results in lower conductivity in the negatively (positively) gated region of the characteristics. However, upon the adsorption of foreign molecules such as polyethyleneimine and aryl diazonium salts, the asymmetry can be reduced or reversed,[9] which indicates that a single transfer curve cannot be used to determine the carrier type unambiguously. In this letter, we report an alternative method to experimentally determine the doped carrier type.

The bottom left panel of Fig. 1 shows a phenomenological doping profile $V_d(x)$ of a graphene channel, where linearly graded doping occurs from the contact edges to a distance $L_d$, and the doping level becomes zero at points distant from the contact. If the channel length $L$ becomes shorter than $2L_d$, the doping effects from two (source and drain) electrodes are overlapped and the central region of the channel is no longer charge neutral, as schematically illustrated in the bottom right panel of Fig. 1. Due to the charge density pinning at the metal contacts, the general V shape of the transfer characteristics is mainly determined by the central region of the channel, where the gate voltage $V_G$ can tune $V_d(x)$ effectively. Thus, $V_G$ which dictates the minimum conductivity (the charge neutrality point or the Dirac point $V_{NP}$) would correspond to $V_d(L/2)$. This scenario shows that $V_{NP}$ shifts to more negative (positive) $V_G$ values as $L$ becomes shorter in the case of electron (hole) doping from metal contacts to a graphene channel. Therefore, the doped carrier type is predicted by examining the direction of $V_{NP}$ shift with the shortening of $L$.

To explore the above concept, graphene FETs were fabricated on a Si substrate covered with a 300-nm-thick thermal oxide layer. Mechanical exfoliation from a graphite crystal (Kaneka, Super Graphite[10]) was used to form graphene sheets. The sheets were determined to be single-layer graphene by means of Raman spectroscopy, and used as formed without further reshaping processes. Conventional lithographic techniques (electron-beam lithography, thermal evaporation, and liftoff) were employed to form metal contacts on the graphene layers. Several metallic electrodes with different inter-electrode spacings were fabricated on the same graphene sheet in order to investigate the $L$-dependent $V_{NP}$ shift using a single graphene sheet. All electrode width along the channel direction was 1.0 μm. Au, Cu and Ag were employed as electrode materials. For Au electrodes, a very thin (<1 nm) adhesion layer of Cr was deposited prior to Au deposition (45 nm in thickness) to strengthen the adhesion of Au to the $SiO_2$ surface. The adhesion layer was sufficiently thin to ensure the growth of small Cr islands on graphene, hence the graphene-metal interface would be predominantly determined by Au-graphene contacts. For Cu and Ag electrodes, a thick (200 nm) cap layer of Au was deposited after Cu or Ag deposition (50 nm in thickness) to avoid the oxidation of Cu or Ag. Graphene channel widths $W$ of Au-, Cu-, and Ag-contacted devices were 2.0, 3.5, and 3.0 μm, respectively. After electrode fabrication, the constructed devices were introduced into a vacuum probe system (base pressure: ca. $10^{-2}$ Pa), and heated at 120 °C for 12 h to remove resist residues. It was confirmed that $V_{NP}$'s of two FETs with the same channel length on the same graphene sheet became almost identical after the heating process. Without the heating process, large inhomogeneity remains, even on the same graphene sheet.[11] All $L$ dependence was measured in the same vacuum after the heating procedure, and using a semiconductor parameter analyzer (Agilent, 4155C) connected to the vacuum probe system. The drain voltage $V_{DS}$ was set to 10 mV.

Figure 2 shows the transfer characteristics of Au-contacted graphene FETs with different channel lengths. The vertical axis represents the total conductivity including contact resistances, which is equal to $I_\mathrm{D}L/(V_\mathrm{DS}W)$. The $V_\mathrm{NP}$ values of these devices were monotonically shifted from −43 to −71 V by the shortening of $L$ from 2.5 to 0.4 μm, which clearly indicates electron doping from Au contacts to graphene. Shifts of the $V_\mathrm{NP}$ values for graphene FETs with Au, Cu, and Ag electrodes are compiled in Figs. 3(a)-3(c), respectively. While a negative $V_\mathrm{NP}$ shift was observed with the shortening of $L$ with Au electrodes, positive shifts were observed with Cu or Ag electrodes, which indicates hole doping from the Cu or Ag contacts to the graphene channels. If the metal-contact doping becomes more distinct by shortening $L$, then the asymmetry of the electron-hole conductivity should also become clearer.[5] Figure 4 shows the $V_\mathrm{G}$ dependence of the ratios of the electron-side to hole-side conductivities. Although the data for Au contacts includes a deviation ($L$=2.5 μm), these plots show that the asymmetry became clearer by shortening $L$.

The respective work functions for Au, Cu, Ag, and graphene are 5.1, 4.65, 4.28,[12] and 4.6 eV.[13] If we adopt the simple idea that the charge transfer between two materials can be determined by comparing the individual material work functions, then electrons (holes) are expected to be doped from Ag (Au or Cu) electrodes. However, the experimental results reported so far are not consistent with this expectation. The type of carriers doped from metal atoms to graphene layers grown on Ni(111) by chemical vapor deposition, where metal atoms are intercalated into the graphene-Ni interface, has been determined by photoemission to be hole, electron, and electron for Au, Cu, and Ag, respectively.[14] On the other hand, transport measurements have shown that Au and Pt (work function: 5.65 eV[12]) atoms deposited onto mechanically exfoliated graphene dope electrons to graphene.[15,16] As such, controversial results have been reported to date.

Differences in experimental conditions, such as the preparation method of graphene and the form of metals, might have affected the extracted carrier types.

The doped carrier types determined in this study for metallic electrodes formed on mechanically exfoliated graphene are also not consistent with the simple expectation. To understand this behavior, modification of the metal work functions by interaction with graphene should be taken into consideration. At metal surfaces, electrons spilling out into the vacuum form a surface electric double layer, and the resultant potential barrier at the surface is known to contribute to the metal work functions.[17] If the vacuum is replaced by graphene, then the spilled electrons are pushed back into the interior of the metal by Pauli repulsion from the p electrons of graphene.[18] The metal work function decreases from its original value as a result of this pushback effect, and could explain the electron doping from Au to graphene if the pushback effect is larger than 0.5 eV. If other types of electric double layers are present at metal-graphene interfaces, then this could result in further modification of the metal work functions. The most popular origin of the additional double layer is chemical interaction between the metal and graphene. A simple index of chemical interaction is the electronegativity of atoms: 2.54, 1.90, 1.93, and 2.55 for Au, Cu, Ag, and C, respectively.[19] If the electronegativity of C atoms is adopted for graphene, then ionic bonds can be expected to form at interfaces between graphene and Cu or Ag, while almost no ionic character can be expected at the interface with Au. The electronegativity of C is larger than those of Cu and Ag; therefore, negative and positive charges may locate at the graphene and metal sides, respectively. The additional electric double layer effectively increases the metal work function, and this could explain the hole doping from Cu or Ag to graphene if the chemical effect is larger than 0.8 eV, which can be estimated by adding the work function difference between Ag and graphene (*ca*. 0.3 eV) to the pushback effect (possibly larger than 0.5 eV).

In conclusion, we have described a method of determining the carrier type doped from metal contacts to graphene. The Dirac points are expected to shift toward more negative (positive) gate voltages with the shortening of the channel length, due to electron (hole) doping from the metal contacts. Experimental examination of the direction of the Dirac point shift indicated that electrons (holes) were doped from Au (Cu or Ag) electrodes. The carrier types determined do not correspond to that expected from a simple comparison of individual work functions. Instead, modified metal work functions, due to interaction with graphene, could correspond to the experimental results.

Acknowledgements  The authors are grateful to M. Murakami and M. Shiraishi for supplying the graphite crystal. This work was supported in part by the Foundation Advanced Technology Institute, Japan.

FIGURE CAPTIONS

Fig. 1. Schematic diagram of the fabricated device structure with several electrodes on a single graphene flake, and phenomenological doping profiles along a graphene channel (channel length, $L$) without application of gate voltage. The vertical axis is a gate-voltage equivalent and positive (negative) values correspond to electron (hole) doping.

Fig. 2. Transfer characteristics of Au-contacted graphene FETs with different channel lengths $L$ (2.5 to 0.4 μm) fabricated on the same graphene flake. The bottom panel is a close-up near the Dirac points showing the gate voltages that yield the minimum drain current. Shortening of $L$ leads to a shift in the Dirac point toward more negative values, which indicates electron doping from Au contacts to the graphene channel.

Fig. 3. Channel length dependence of the Dirac point for (a) Au-, (b) Cu-, and (c) Ag-contacted graphene FETs. The values for the Au-contacted device were taken from Fig. 2. The negative (positive) shift of the Dirac point by shortening of the channel length corresponds to electron (hole) doping from Au (Cu and Ag) electrodes.

Fig. 4. Gate voltage dependence of the ratio of the electron-side to hole-side conductivities for (a) Au-, (b) Cu-, and (c) Ag-contacted graphene FETs. The asymmetry of the electron-hole conductivity was generally found to be enhanced by shortening the channel length.

Fig. 1.

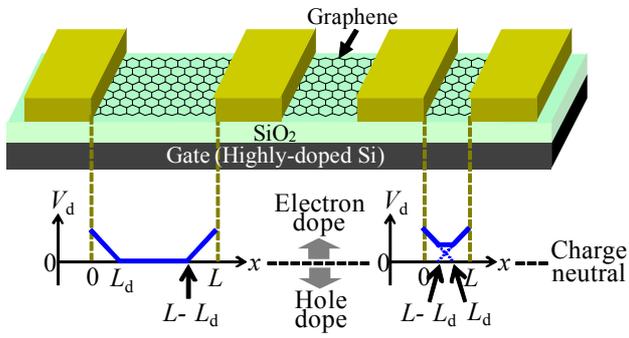

Fig. 2.

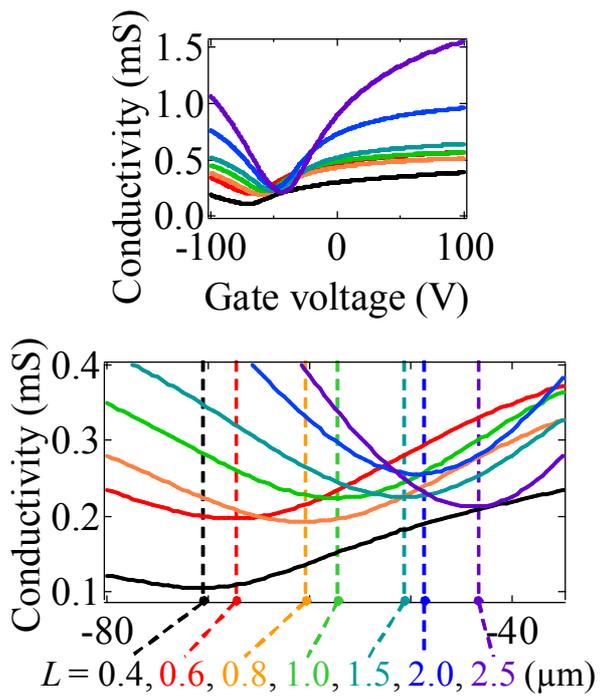

Fig. 3.

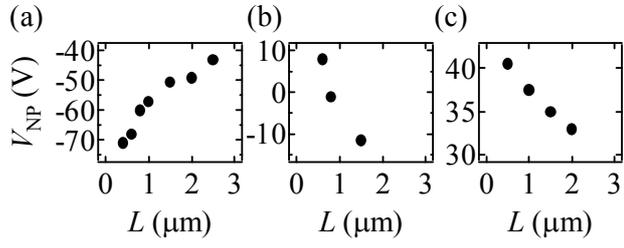

Fig. 4.

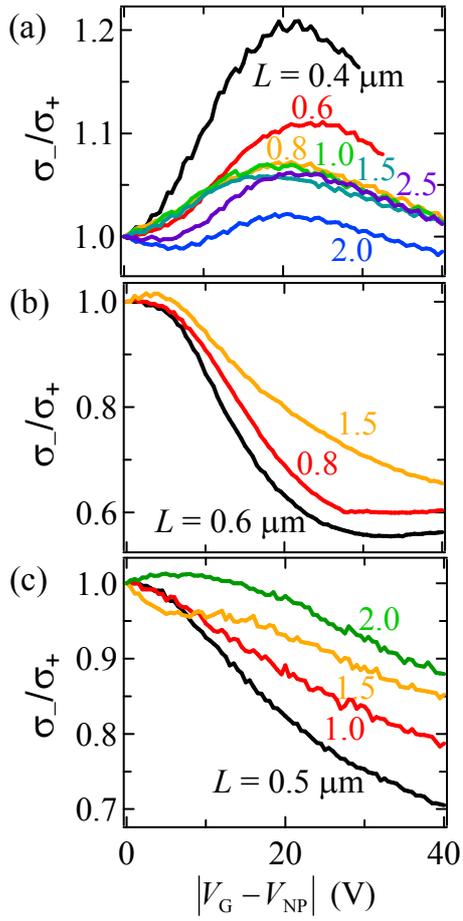

12